\documentclass[a4paper,11pt]{article}
\pdfoutput=1 

\usepackage{jinstpub} 
\usepackage{subfigure}

\title{\boldmath Photon detection system for ProtoDUNE dual phase}

\author[a]{C. Cuesta\note{Corresponding author.}}

\affiliation[a]{Centro de Investigaciones Energ$´{e}$ticas, Medioambientales y Tecnol$´{o}$gicas,
CIEMAT, \\ 28040, Madrid, Spain}

\emailAdd{clara.cuesta@ciemat.es}

\abstract{The Deep Underground Neutrino Experiment (DUNE) is a 40-kton underground liquid argon time-projection-chamber (LAr TPC) detector, for long-baseline neutrino oscillation studies, and for neutrino astrophysics and nucleon decay searches. Photon detector systems embedded within the LAr TPC add precise timing capabilities for non-beam events. The ProtoDUNE dual phase detector will consist of a 6$\times$6$\times$6\,m$^{3}$ liquid argon time-projection chamber placed at CERN and the light readout will be formed by 8-inch cryogenic photomultipliers from Hamamatsu. The characterization of the 36 photomultipliers, the base design, and the light calibration system are described. In addition, preliminary results from a  3$\times$1$\times$1\,m$^{3}$ LAr double phase detector operating at CERN are presented.}

\keywords{Noble liquid detectors (double-phase); photon detectors (photomultipliers); neutrino detectors;}

\arxivnumber{1711.08307} 

\collaboration[c]{on behalf of DUNE collaboration}

\proceeding{LIDINE 2017: Light Detection in Noble Elements\\
  22-24$^{th}$ September 2017\\
  SLAC National Accelerator Laboratory}

\begin{document}
\maketitle
\flushbottom

\section{ProtoDUNE dual phase}
\label{sec:protoDUNE}

The DUNE experiment aims to address key questions in neutrino physics and astroparticle physics~ \cite{duneCDRv2}. It includes precision measurements of the parameters that govern neutrino oscillations with the goal of measuring the CP violating phase and the neutrino mass ordering, nucleon decay searches, and detection and measurement of the electron neutrino flux from a core-collapse supernova within our galaxy. DUNE will consist of a near detector placed at Fermilab close to the muon neutrino beam generated at the so-called Long-Baseline Neutrino Facility (LBNF), and four 10\,kt fiducial mass LAr TPCs as far detector at 1300\,km from Fermilab in the Sanford Underground Research Facility (SURF) at 4300\,m.w.e. depth~\cite{duneCDRv4}. 

In order to gain experience in building and operating such large-scale LAr detectors, an R\&D programme is currently underway at CERN~\cite{ProtoDUNEs}. Such programme will operate two prototypes in a dedicated test beam by 2018, with the specific aim of testing the prototypes design, assembly, and installation procedures, the detectors operations, as well as data acquisition, storage, processing, and analysis under beam conditions and with cosmic ray data. The two prototypes will both employ LAr TPCs as detection technology, with one prototype only using liquid argon, hereby called ProtoDUNE Single-Phase, and the other using argon in both its gaseous and liquid state, thus the name ProtoDUNE Dual-Phase (DP). Both detectors will have similar sizes. In particular, ProtoDUNE-DP~\cite{wa105}, also known as WA105, will have an active volume of 6$\times$6$\times$6 m$^{3}$ and an active mass of 300\,t. In ProtoDUNE-DP the charge read-out occurs in the gas phase, where the charge is extracted, amplified, and detected in gaseous argon above the liquid surface allowing a finer readout pitch, a lower energy threshold, and better pattern reconstruction of the events. The light signal is used as trigger for non-beam events, and t$_{0}$ for both beam and non-beam events (cosmic background rejection), and there is also a possibility to perform calorimetric measurements and particle identification, as it is done in the dark matter experiments~\cite{darkside,ardm}. The photon detection system of ProtoDUNE-DP is formed by 36 8-inch cryogenic photomultipliers (PMTs) placed at the bottom of the detector. As wavelength-shifter, tetraphenyl butadiene (TPB) is coated directly on the PMTs. Each photomultiplier has a voltage divider base soldered to it, whose design is described in section~\ref{sec:PMT}.1. In addition, a light calibration system checks the optimum operation of the PMTs, see section~\ref{sec:LCS}, and an external DAQ system records the light signal.


\section{The 3$\times$1$\times$1\,m$^{3}$ LAr TPC}
\label{sec:311}

\begin {figure}[ht]
\includegraphics[width=0.65\textwidth]{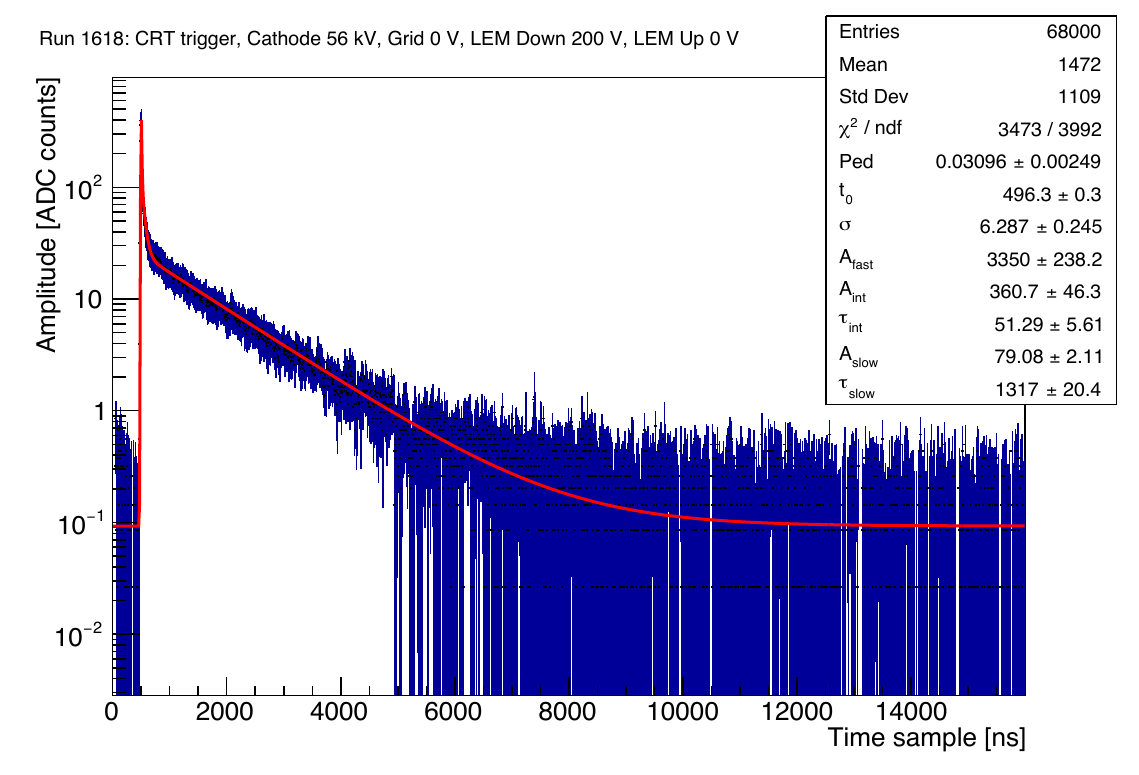}
\centering \caption{\it Averaged waveform of the S1 light signal taken with one PMT from the 3$\times$1$\times$1\,m$^{3}$ LAr DP TPC, fitted with a function (red line) that is the sum of a Gaussian, parametrized by t$_{0}$ and $\sigma$, and two exponential functions, with decay time constants $\tau_{fast}$ and $\tau_{slow}$, and normalization factors $A_{fast}$ and $A_{slow}$.}
\label{fig:311light}
\end {figure}

Following a staged approach, prior to the ProtoDUNE-DP construction, the WA105-3$\times$1$\times$1\,m$^{3}$ dual phase LAr-TPC demonstrator was assembled at CERN~\cite{311}. Its construction and operation aims to test scalable solutions for the crucial aspects of this technology: ultra high argon purity in non-evacuable tank, large area dual phase charge readout system in several square meter scale, and accessible cold front-end electronics. Five PMTs coated with the wavelength shifter, TPB, are fixed under the cathode. They are sensitive to the 128 nm scintillation light from the argon scintillation and provide the reference time for the drift as well as the trigger. The detector is taking cosmic ray data since summer 2017, and the results are being analysed. Different analyses are on-going using the light response. An example of an averaged waveform fitted to a fast and a slow scintillation components~\cite{aprile} is shown in Figure~\ref{fig:311light}. As the slow scintillation component can be monitored run by run on every PMT, the LAr purity is being studied. The relation between the S1 signal coming from interactions in the liquid phase and the S2 signal due to the amplification gaseous phase is being studied to determine the best operating conditions of the photon detection system. Also, the response to muon interactions is under study, the light and charge signals are being correlated, and a comparison of data to simulation is under preparation.

\section{Photomultiplier characterization}
\label{sec:PMT}

The photon detection system of ProtoDUNE-DP uses 8-inch Hamamatsu R5912-20Mod cryogenic PMTs~\cite{ham}. R5912-20Mod is a 14-stage high gain PMT with a bi-alkali photocathode. The PMT has been adequately modified to compensate the gain diminution expected in a cryogenic environment, so 14 dynodes are used instead of 10. In addition, a thin platinum layer has been added between the photocathode and the borosilicate glass envelope to preserve the conductance of the photocathode at low temperature increasing the quantum efficiency. A PMT mount has been designed, manufactured, and assembled at CIEMAT.  The support frame structure has been made of 304\,L Stainless steel and Nylon 6.6 pieces assembled by A4 stainless steel screws that minimize the mass while ensuring the PMT support to the cryostat membrane. The PMTs will be placed at the ProtoDUNE-DP membrane corrugation squares and the base will be glued to the membrane.

\subsection{PMT testing setup}
\label{sec3}

\begin {figure}[ht]
\subfigure[]{\includegraphics[width=0.6\textwidth]{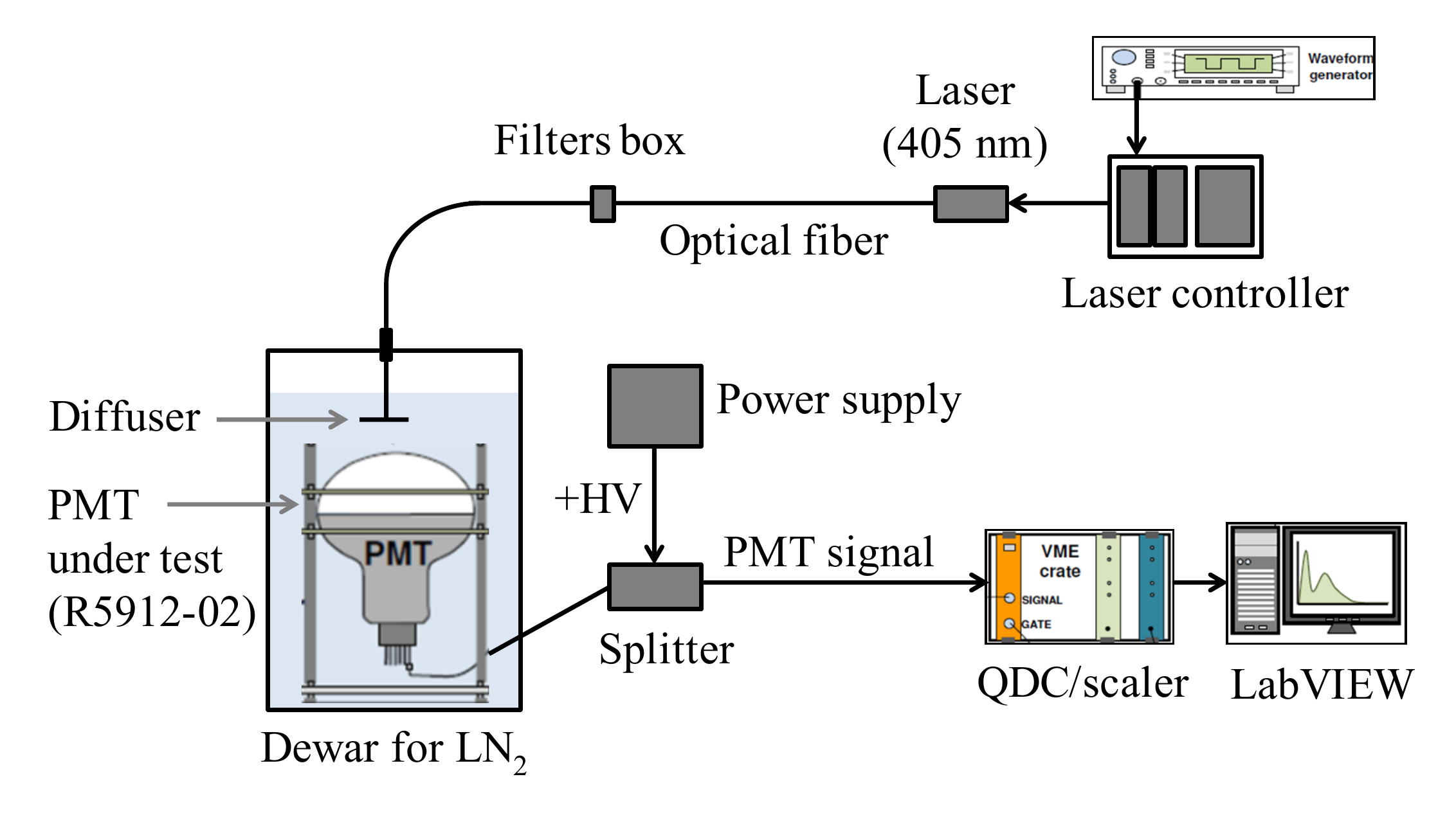}}
\subfigure[]{\includegraphics[width=0.22\textwidth]{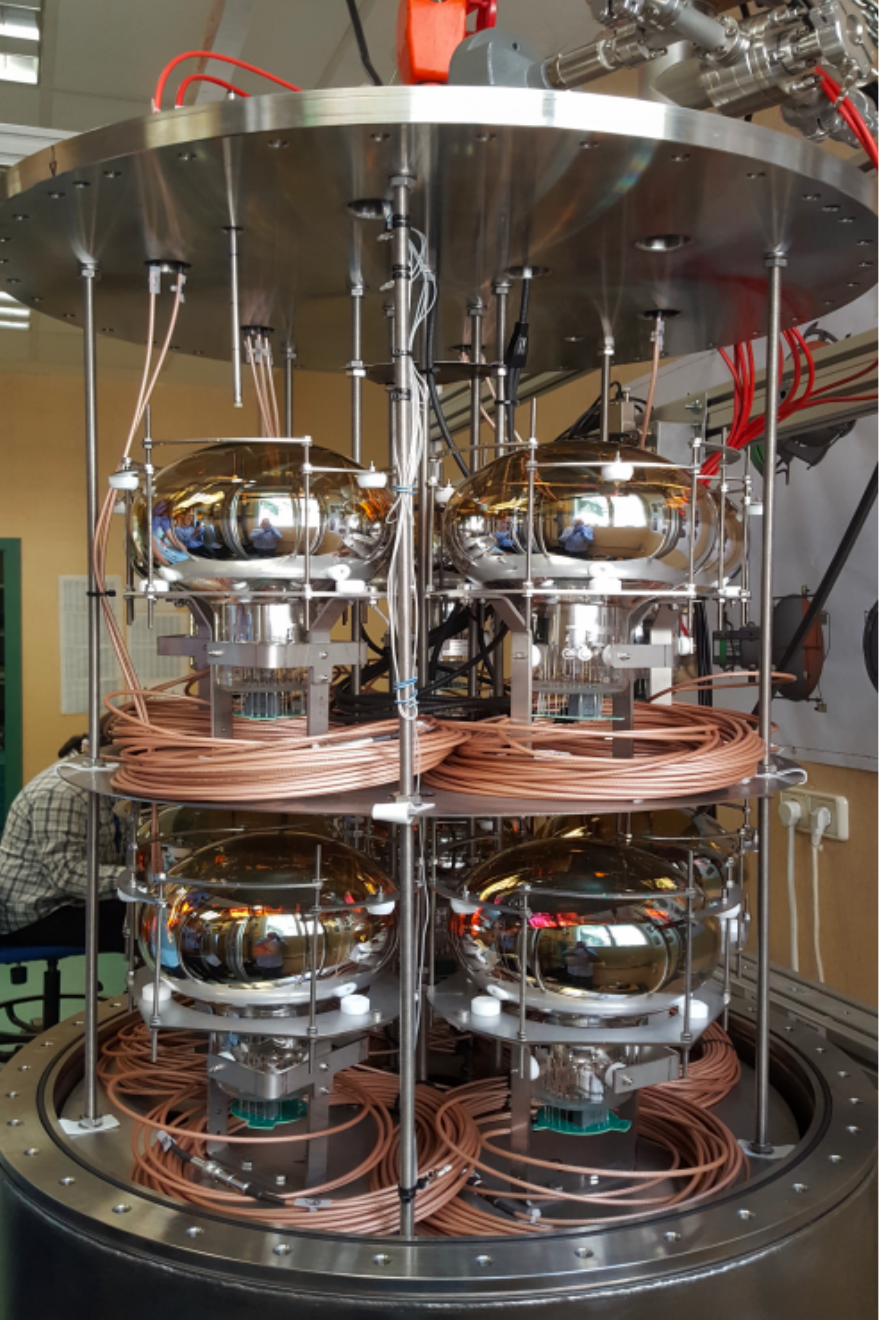}}
\centering \caption{\it (a) Schematic drawing of the PMT testing experimental setup. (b) Picture of 10 PMTs installed in the testing dewar.}
\label{fig:setup}
\end {figure}

A dedicated test bench was designed at CIEMAT for the PMT characterization. A schematic drawing is shown in Figure~\ref{fig:setup}.a. The characterization measurements were performed to 10 PMTs at the same time inside a 300\,L vessel at room temperature (RT) and filled with liquid nitrogen for the cryogenic temperature (CT) tests. Figure~\ref{fig:setup}.b shows a picture of the vessel with 10 PMTs being installed. A laser with 405\,nm wavelength is used to provide a known amount of light to the PMTs operating with a controller. The signal from the PMTs is read through the splitter that separates the HV and the PMT signal, then using a QDC (gain measurements) or a scaler (DC measurements), and LabVIEW software.

\subsection{PMT base design}
\label{secBase}
Different voltage divider configurations were tested in order to choose the optimal base design to be used in ProtoDUNE-DP. In the so-called positive base (PB), the high voltage (HV) is applied at the anode, whereas in the negative base (NB) it is applied at the cathode. The PB is configured using only one cable as it is possible to apply at the same time the HV, and read the charge signal at the PMT anode with the photocathode being grounded. Then, a splitter circuit separates the HV and the charge signal out of the cryostat. This modification affects the real voltage value that is sent to the PB which is expected to be a 6\% lower than the one read on the power supply system itself. On the other hand, in the NB negative HV is supplied through the PMT cathode and the output charge signals are carried out/in separately, using two cables. The PB design was selected after some validation measurements because it reduces the total number of cables, and then the cost. Once the PB was selected as final design, all the PMT bases were mounted, cleaned and tested at CIEMAT. Two tests were performed before soldering to PMTs: the total resistance is 13.44 M$\Omega$ (with the tolerance margin 0.1\%), and they were tested at 2000 V in Ar gas to verify the absence of sparks. Then, all the PMT bases were soldered to the PMTs.


\subsection{Measurements and results}
\label{sec4}

In order to study the performance of the PMTs, different tests are carried out at RT and CT. First, dark current is studied measuring the event rate in total darkness, see section~\ref{sec4.1}. Then, the PMT gain is quantified  compared to the one provided by Hamamatsu, as explained in section~\ref{sec4.2}. Further tests to study the PMT response with the incident light are on-going. In these studies, the amount and frequency of the light sent to the PMT surface is increased until the PMT saturates.

\subsubsection{Dark Current}
\label{sec4.1}

\begin {figure}[ht]
\subfigure[]{\includegraphics[width=0.4\textwidth]{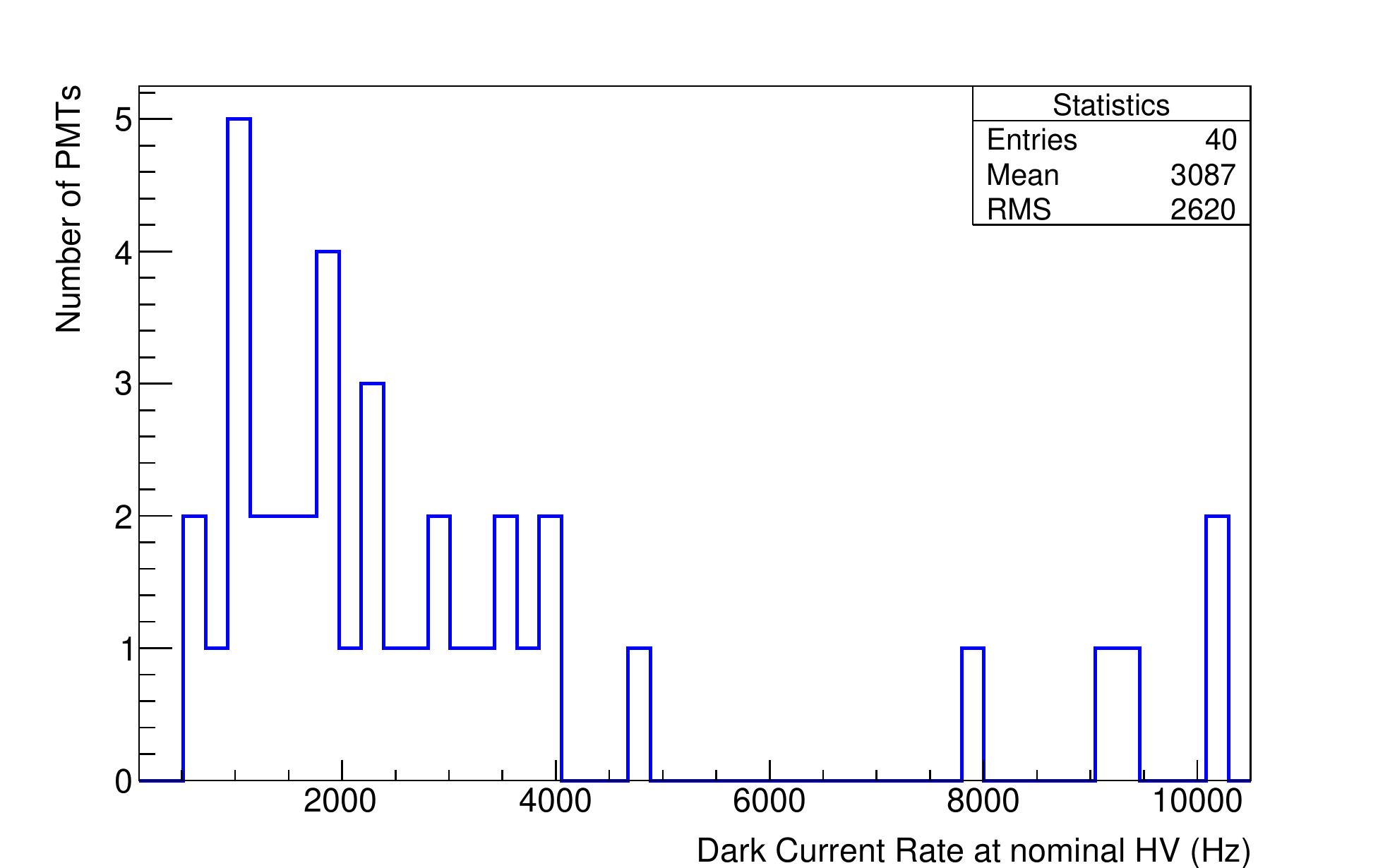}}
\subfigure[]{\includegraphics[width=0.4\textwidth]{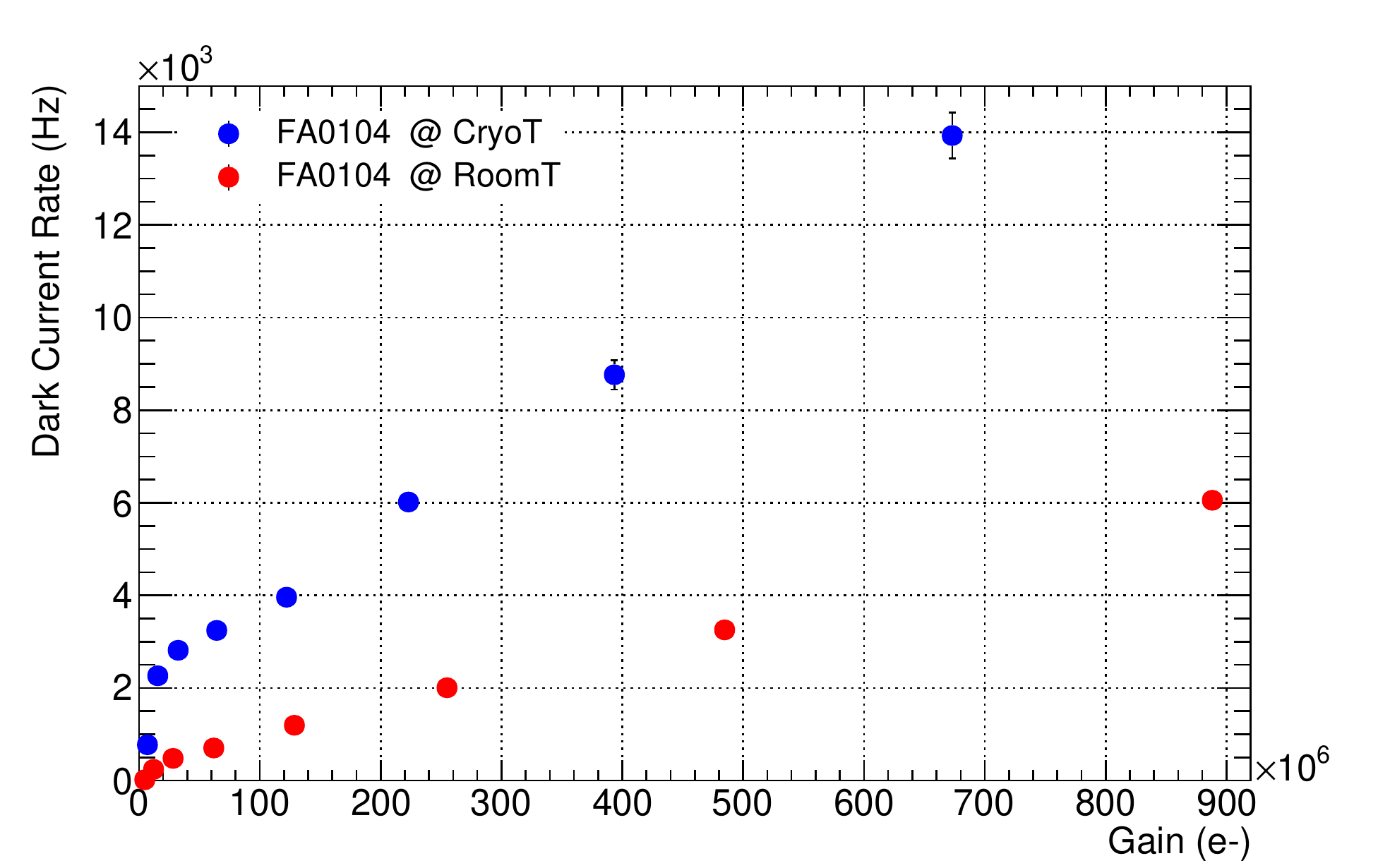}}
\centering \caption{\it (a) Dark current at RT for the Hamamatsu R5912-20Mod cryogenic PMTs to be used in ProtoDUNE-DP for a 10$^{9}$e$^{-}$ gain . (b) Comparison of the DC vs gain for a PMT at RT and CT.}
\label{fig:dc40}
\end {figure}

The dark current (DC) is a current measured at the anode of the PMT in absence of incident light. It depends on the cathode composition and its behaviour as a function of the voltage applied varies with the gain. 
The DC rate corresponds to the number of spontaneous SPE charge signals that pass a threshold defined as half the mean SPE amplitude at 10$^{7}$e$^{-}$ gain to ensure trigger over the noise level and that SPE triggers at any PMT gain. It is worth mentioning that these measurements were done after leaving the PMT in darkness for at least two days. DC was measured at RT for a 10$^{9}$e$^{-}$ gain for all the PMTs, see Figure~\ref{fig:dc40}.a, with similar results as those provided by Hamamatsu. At CT, DC was measured too, and the comparison of the DC behaviour with gain for a PMT is shown in Figure~\ref{fig:dc40}.b. When the PMT works at CT, the causes that generate the DC are different from at RT, but not all the causes are completely known yet \cite{Nikkel,meyer}. When the PMT is cooled down, the thermoionic DC rate is expected to decrease with the temperature according to the Richardson's law. Despite that, it has been observed that an experimental deviation from this prediction exists, due to non-thermal contributions~\cite{Nikkel,uBoone}. 

\subsubsection{Gain}
\label{sec4.2}

\begin {figure}[ht]
\subfigure[]{\includegraphics[width=0.4\textwidth]{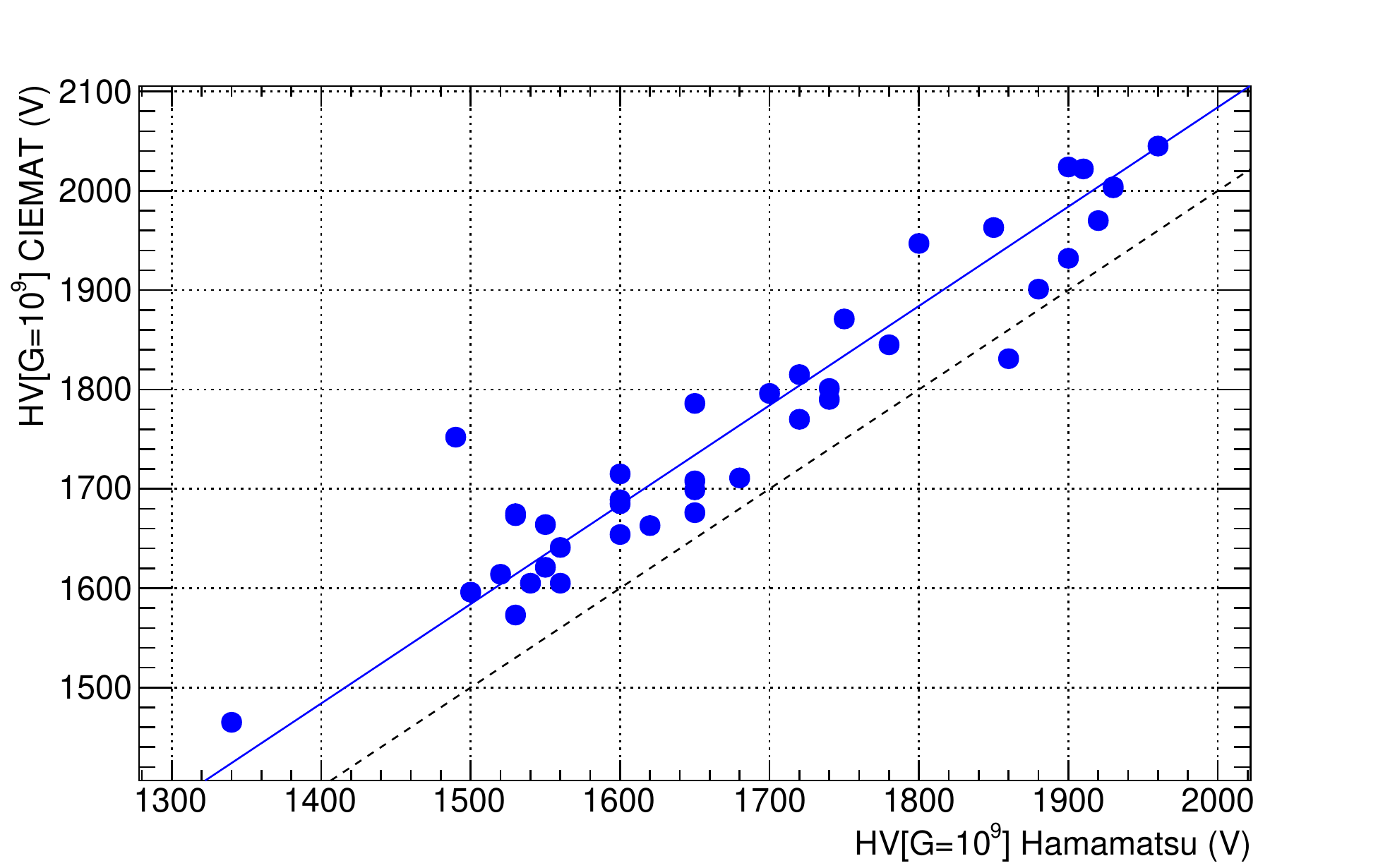}}
\subfigure[]{\includegraphics[width=0.4\textwidth]{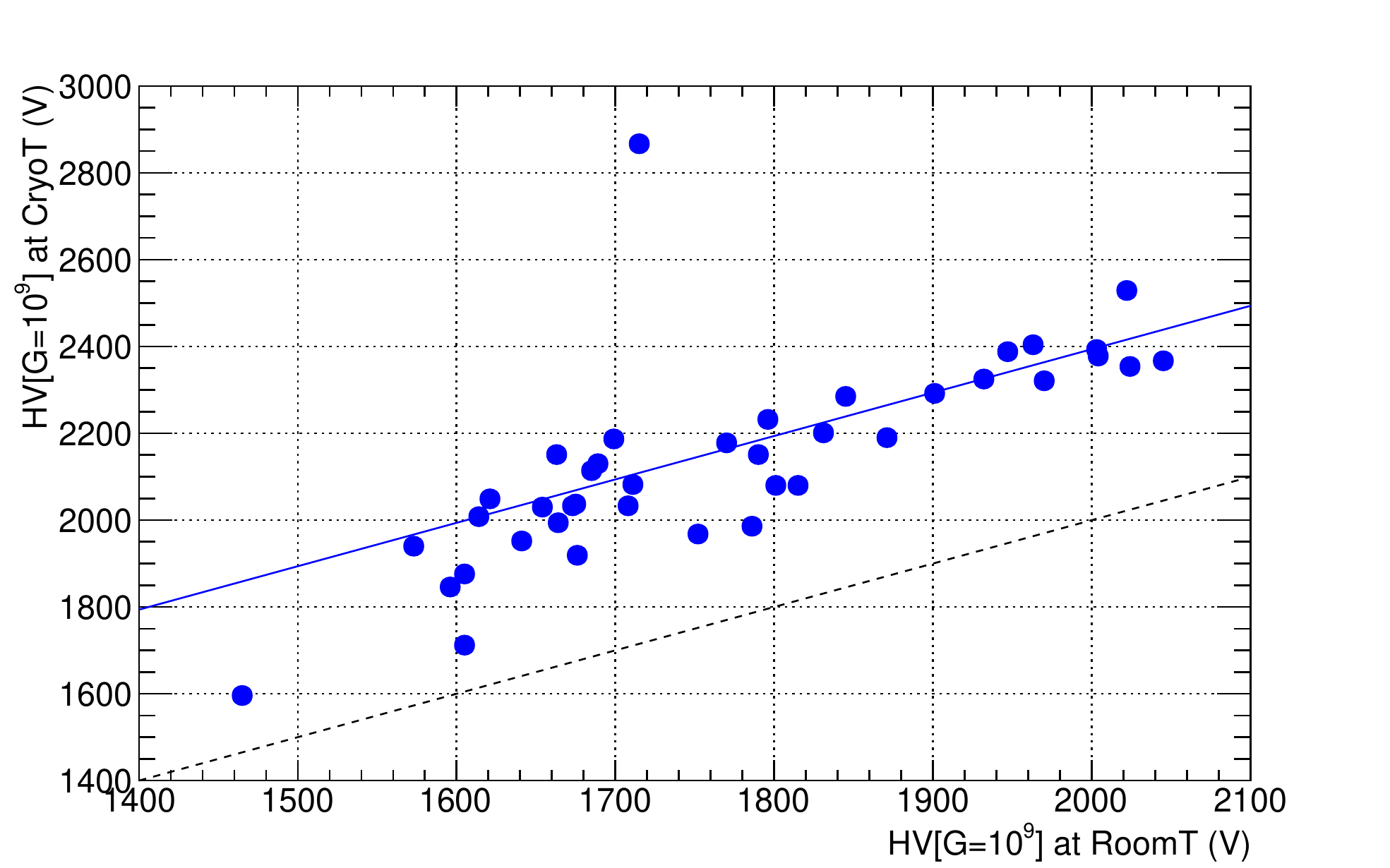}}
\centering \caption{\it HV applied to obtain a gain of 10$^{9}$e$^{-}$ measured at CIEMAT compared to the values provided by Hamamatsu at RT (a), and CIEMAT results at CT compared to RT (b).}
\label{fig:grtct}
\end {figure}

The gain of a PMT can be extracted from the analysis of the PMT charge spectrum corresponding to a weak pulsed light source, the so-called single photo-electron spectrum. The gain has been measured for all PMTs at RT and CT, and compared to the values provided by Hamamatsu. Similar values to the ones provided by Hamamatsu are measured at RT, see Figure~\ref{fig:grtct}, with an average of 83 V difference as expected by the PB design. At CT, Hamamatsu values are not provided and CIEMAT measurements show that higher voltages need to be applied to obtain the same gain. For instance, to obtain a gain of 10$^{9}$e$^{-}$, at CT an average of 371 V (21\%) more than at RT need to be applied. It is worth highlighting that PMT FA143 showed a significant lower gain at CT without being correlated to any other unexpected behavior.

\section{Light calibration system}
\label{sec:LCS}

A light calibration system integrated in ProtoDUNE-DP detector has been designed to monitor the calibration of the 36 PMTs installed inside the LAr volume. The goal is to determine the PMT gain and study the PMT stability. An optical fiber will be installed at each PMT in order to provide a configurable amount of light. The calibration light will be provided by a blue LED of 470\,nm using a Kapuschinski circuit as LED driver which reduces significantly the cost of using a laser. There will be one LED connected to one fiber going to one optical feedthrough, so there will be six LEDs in total placed in a hexagonal geometry. The direct light will go to the fiber, and the stray light to a SiPM used as reference sensor, being a single reference sensor in the center. Then, there will be six fibers of 22.5-m inside the cryostat. Each one of these fibers will be attached to a 1-to-7 fiber bundle, so that a fiber is finally installed at each PMT. Several tests to quantify the light losses of this design were performed successfully.

\acknowledgments
This project has received funding from the European Unions Horizon~2020 Research and Innovation programme under Grant Agreement no.~654168 and from the Spanish Ministerio de Econom$´{i}$a y Competitividad under Grant no.~FPA2016-77347-02-1.


\end{document}